\documentclass[12pt]{article}

\usepackage{epsfig}

\usepackage{times}

\title{{\it In-situ} Laser Microprocessing at the Quantum Level} 

\author
{Armando Rastelli,$^{1\ast}$ Ata Ulhaq,$^{1}$ Suwit Kiravittaya$^{1}$, Lijuan Wang$^{1}$,\\ Artur Zrenner$^{2}$, Oliver G. Schmidt$^{1}$\\
\\
\normalsize{$^{1}$Max-Planck-Institut f\"ur Festk\"orperforschung, Heisenbergstr.\ 1, D-70569 Stuttgart, Germany}\\
\normalsize{$^{2}$Universit\"at Paderborn, Experimentalphysik, Warburgerstr.\ 100,
D-33098 Paderborn, Germany}\\
\\
\normalsize{$^\ast$To whom correspondence should be addressed; E-mail:  a.rastelli@fkf.mpg.de.}
}

\begin{document} 




\maketitle


\begin{abstract}
One of the biggest challenges of nanotechnology is the fabrication of nano-objects with perfectly controlled properties. Here we employ a focused laser beam both to characterize and to {\it in-situ} modify single semiconductor structures by heating them from cryogenic to high temperatures. 
The heat treatment allows us to blue-shift, in a broad range and with resolution-limited accuracy, the quantized energy levels of light and charge carriers confined in optical microcavities and self-assembled quantum dots (QDs). We demonstrate the approach by tuning an optical mode into resonance with the emission of a single QD and by bringing different QDs in mutual resonance. 
This processing method may open the way to a full control of nanostructures at the quantum level.
\end{abstract}

In material science lasers are widely employed both for characterization and for processing/machining. 
In the field of semiconductors, the local heat produced by a laser source has been used, e.g., to anneal defects created by implantation~\cite{poate82}, to locally crystallize amorphous semiconductors~\cite{poate82,allmen95,chimmalgi05}, to fabricate lateral transistors by local interdiffusion of doping atoms~\cite{baumgartner94}, to locally modify the wall structure of rolled-up nanotubes~\cite{deneke04}, to produce nanostructures by local intermixing of quantum wells (QWs) with the surrounding material~\cite{brunner92-1,nebel97,zaitsev01}, and to selectively blue-shift the emission of quantum dot (QD) arrays~\cite{dubowski00,djie06}.
Spatial resolutions much beyond the optical diffraction limit can be achieved~\cite{brunner92-1,chimmalgi05-1}. These examples give a flavour of the wide range of applications accessible by laser processing. 
In most of the studies presented so far the characterization of the fabricated/processed structures was accomplished {\it ex-situ}, with a setup independent from that used for the processing.

Here we use the same focused laser beam both for characterization (at low power) and for post-fabrication processing (at high power) of single micro- and nano-structures, which are placed in the cryostat of a micro-photoluminescence ($\mu$-PL) setup. We thus refer to this approach as an {\it in-situ} process, analogous to scanning probe microscopy, where a tip can be used both to measure and to modify the structure under investigation~\cite{crommie93}. 
We consider semiconductor optical microcavities~\cite{vahala03} and self-assembled quantum dots~\cite{michler03} (QDs), which confine light and charge carriers in all directions, respectively. The confinement results in the appearance of discrete energy states which are observed as sharp lines (with linewidths in the $\mu$eV-meV range) in the emission spectra.
The energies of the confined states in nominally identical structures are usually different, because of their high sensitivity to the structural parameters and their unavoidable fluctuations during fabrication. This is a general problem encountered in nanotechnology and is particularly relevant for objects obtained by bottom-up (or self-assembly) techniques. For self-assembled QDs, a fabrication or post-fabrication processing technique able to yield QDs with deterministic emission energies is still sought for~\cite{krauss05}.
We show that by {\it in-situ} laser processing (ILP) we can engineer, within a broad range and resolution-limited accuracy, the energy states of both light and charge carriers confined in single microdisk resonators and QDs. This allows us to controllably blue-shift the optical modes and to produce arrays of perfectly resonant QDs.

The samples studied here are fabricated by solid-source molecular beam epitaxy (MBE) on GaAs(001) substrates and processed to obtain microdisk resonators or square-shaped mesas on top of thin posts (Fig.~1A). The active region contains self-assembled In(Ga)As QDs with low surface density ($<$1~$\mu$m$^{-2}$, see inset of Fig.~1A) on top of a thin wetting layer (WL). A SiO$_x$ layer is deposited on top of the structures to partially limit arsenic desorption during the laser-heating process~\cite{dubowski00}.
The samples, placed inside a He-flow cryostat, are investigated and laser-processed in a standard $\mu$-PL setup equipped with a continuous-wave laser with emission wavelength of 532~nm. The laser is focused to a spot with $\sim$1.5~$\mu$m diameter.

Figure 1A schematizes the {\it in-situ} laser processing (ILP) approach, which consists in using the focused laser beam both as a probe and as a tool to engineer the microdisk and QD emission properties. 
At low laser powers (nW-$\mu$W range) we measure the QD and cavity-mode emission, while at relatively high powers (mW range) we heat the structures. Figures 1B and 1C show an example of the evolution of the PL from a microdisk with 4-$\mu$m diameter while the laser power is ramped from 150 nW up to 3.8 mW in 100 s. At low power the microdisk temperature $T_d$ is the same as the temperature of the substrate (5~K). With increasing power the emission of the WL and GaAs matrix are observed to red-shift. The same occurs for the modes (sharp lines, see also inset of Fig. 1C), although the amount of the shift is much smaller. This behaviour indicates a progressive {\it heating} of the microdisk. 
The red-shift of the GaAs and WL emission and that of the modes are due mainly to band-gap shrinkage and increase of the effective refractive index of the material with increasing temperature, respectively.
By knowing the temperature dependence of the optical properties of GaAs~\cite{brozel96} we can estimate $T_d$ as a function of the laser power. Since the GaAs and WL peak broaden with increasing temperature (see spectra in Fig. 1C), we extract the temperature values shown on the right axis of Fig.~1B from the mode positions, which we use as ``local thermometers''. (See Ref.~\cite{brunner92-1} for a different method to determine the local temperature during laser heating.) The displayed values indicate that it is possible to heat a disk from cryogenic to elevated temperatures with moderate laser powers.

To get a deeper insight in the laser-heating process, we model the microdisk as a 4~$\mu$m-diameter disk of GaAs placed on top of a post with truncated cone shape. We assume the laser to have a Gaussian beam profile with full-width-at-half-maximum of 1.5~$\mu$m and to be centered above the microdisk center. The power absorbed in the disk represents the heat source while the substrate is the thermal bath kept at cryogenic temperatures (5~K). The temperature profile calculated by solving the heat conduction equation by means of a finite element method is graphically displayed in Fig. 1D for a laser power of 4~mW. With this geometry the temperature $T_d$ is rather homogeneous across the disk area, with a slight drop on top of the post. The calculation also suggests that $T_d$ reaches a stable value within a few $\mu$s after the ``switching on'' of the heat source. For the experiments considered here, where the ramping of the power is achieved in about 1~s, the heating and cooling can thus be considered instantaneous. Since the post is the channel through which the heat flows towards the substrate, its size critically affects the value of $T_d$ reachable for a given laser power. This is confirmed by the experimental observation that disks with relatively large posts require several tens of mW to be appreciably heated.

We now focus on the effect produced by the heating process, first on the optical modes and then on the QD emission. 
It is known that residual gases present in the cryostat are cryo-gettered during cooling. Such an adsorption process alters the surface of the microcavity and thus modifies the energy of the confined modes~\cite{strauf06}. 
By heating a single microdisk with the laser we can therefore aim at a controllable desorption of the gettered material, and a consequent blue-shift of the modes. Since the disk temperature goes back to 
5~K as soon as the laser power is decreased, we can check the effect of the desorption after each heating step. This is illustrated in the left panel of Fig.~2A, where a disk with 2.8-$\mu$m diameter is heated several times at the indicated laser powers for periods of 15~s. 
The displayed modes blue-shift after each heating step with no significant deterioration of the mode quality factor $Q$. For the broader peak, which we assign to a transverse-magnetic mode (TM)~\cite{frateschi96}, $Q$ remains in fact constant at about 1700, while for the sharper transverse-electric (TE) mode we observe a slight decrease of $Q$ from 13600 to 10400. The temperature at which appreciable shifts are  first observed is about 70~K, compatible with the onset of desorption of nitrogen and oxygen. Other materials may desorb at higher temperatures. The spectra on the left panel of Fig.~2A also show that different modes shift at different rate. 
The TM mode shifts faster than the TE mode and moves from the low to the high energy side of the TE mode. In order to understand this behavior, which we observe for several disks, we perform a calculation of the mode positions for a microdisk with 2.8~$\mu$m diameter. A symmetric slab with GaAs waveguiding material is used to calculate the effective refractive indices. The energy dispersion relation of the refractive index is also considered and the mode equations are solved consistently, in the whispering gallery mode approximation~\cite{frateschi96}. The best fit of the mode positions prior to processing is obtained for a disk thickness $h$ of 254~nm. 
The relative shifts of TE and TM modes can be reproduced only if we assume that $h$ decreases after each heating step (see right panel of Fig.~2A). By changing the disk diameter we observe in fact that both modes should shift at the same rate. Therefore, this calculation suggests that the desorption occurring during heating mainly affects the disk thickness, consistent with the high ratio between the area of the lower and sidewall disk surfaces. (The top surface is ``passivated'' by the SiO$_x$ cap).

Since the heating steps can be made arbitrarily small by properly tuning the heating time and laser power, a deterministic and fine mode tuning of a single microcavity mode can be achieved by ILP. 
With respect to other {\it ex-situ} approaches reported recently~\cite{badolato05,white05}, the ILP method has the advantage of finesse and immediate feed-back.
The ability of controlling the mode energies is particulary attractive for solid-state-based quantum electrodynamics experiments~\cite{vahala03,yoshie04,reithmaier04,peter05}.
In order to observe coupling phenomena between an optical mode and a single quantum emitter such as a quantized state of a QD, a spatial and spectral overlap have to be achieved~\cite{badolato05}. 
The energies are usually tuned into resonance by changing the sample temperature and exploiting the different temperature dependence of mode and QD emission energies~\cite{badolato05,yoshie04,reithmaier04,peter05,michler00}. This approach is limited in range and has the disadvantage of a degradation of the QD emission at relatively high temperatures due to phonon interaction.
By means of ILP, a certain mode can be shifted to the desired position and then the experiment can be performed at low temperature. 
Figure 2B illustrates the controlled tuning of a mode into (and off) resonance with the negative trion X$^-$ of a single QD. (The assignment follows from power and space-dependent measurements not shown here and is compatible to previous reports for QDs grown under similar conditions~\cite{badolato05,stinaff06}).
When the mode is brought into resonance with the X$^-$ line we observe an enhancement of the emission, which may be  attributed to weak coupling~\cite{michler00}. Since our QDs are randomly positioned inside the microcavity, coupling is not guaranteed even when the energies are perfectly in resonance. Approaches to address this problem have already been demonstrated~\cite{badolato05,xie05}.
While we have considered here only microdisk cavities, ILP is expected to produce similar results also for other cavity geometries such as photonic crystal membranes. This would allow for a fine tuning complementary to the digital etching approach presented in Ref.~\cite{badolato05}.

In Fig.~2B the QD emission remains stable throughout the experiment. By further increasing the power, also the QD emission is observed to blue-shift (not shown). This is expected, since at high enough temperatures the In atoms contained in the QDs and the Ga atoms in the surrounding matrix start to intermix. Bulk-intermixing has an activation energy of about 3.5~eV~\cite{leon97}, i.e. it becomes significant at temperatures of the order of 1000~K for the heating times used here.
Following the same method described above for the mode shift, we can now aim at an {\it in-situ} tuning of the energy positions of single QD emission lines. The emission of self-assembled QDs differs from QD to QD leading to inhomogeneously broadened emission spectra when a large number of QDs is considered. By tuning the growth parameters the inhomogeneous broadening can be substantially reduced, but it has become clear that an array of perfectly resonant QDs cannot be fabricated. Post-processing tuning seems at present the only viable path to obtain QDs with identical emission (within the typical transition linewidth of some $\mu$eV) and to achieve control at the quantum level~\cite{krauss05}.
Figure 3A demonstrates that ILP can indeed be used to bring into resonance spatially separated QDs. For this experiment we consider single QDs located in different mesa structures and we take the positively charged trion X$^+_T$ of a quantum dot, labeled as QDT, as our target. By laser heating for 7~s at increasingly high powers (up to a few mW) we gradually blue-shift the emission of two QDs, QD1 and QD2, until their X$^+$ transitions, X$^+_1$ and X$^+_2$, reach the same energy as X$^+_T$. At the beginning of the ILP we perform relatively large steps, while at the end, when X$^+_{1/2}$ approach X$^+_T$ we perform smaller steps. Figure 3B shows a summary of the X$^+_1$ and X$^+_2$ positions as a function of heating step. We can illustrate the mechanism leading to the blue-shift by assuming that the different QDs have the same homogeneous composition, but different sizes prior to processing. 
The interdiffusion occurring during the heat treatment smoothes the interfaces between QD and surrounding barrier, leading to shallower confinement potentials for electrons and holes, as depicted in the insets of Fig.~3B. In reality different QDs are characterized by different atomic arrangements~\cite{offermans05}, since all the processes taking place during InAs deposition and subsequent GaAs-overgrowth can be described by statistics. This renders the structure, and hence the optical spectra of each QD unique (note, e.g., the X$^0$-X$^+$ separation in the 3 QDs shown in Fig.~3A prior to ILP). While we have focused on the positions of the X$^+$ lines during ILP, we find that also the X$^0$-X$^+$ separations become more homogeneous, although not identical, when the X$^+$ lines are tuned into resonance. This unexpected result will be subject of further investigations.
From Fig.~3 we observe that ILP allows the QD emission to be shifted in a broad range (about 15~meV for QD2), comparable with typical inhomogeneous broadening values. The emission lines remain resolution-limited throughout the process, suggesting that no significant damage is produced by the ILP. The processing is at the same time at least as accurate as the resolution of our spectrometer (70~$\mu$eV). This is illustrated more in detail in the inset of Fig. 3A, which displays a series of spectra corresponding to the final ILP steps for QD1. Since the heating steps can be made arbitrarily small, we are confident that lines of different QDs can be brought into perfect resonance, i.e., to coincide within their intrinsic linewidths.
We can thus conclude that ILP can be used to fabricate arrays of perfectly resonant quantum emitters. 

Finally we want to discuss the possible future extentions of the ILP method. For what concerns the control of cavity modes, the tunability range may be probably increased by intentionally coating the cavity surface with materials with known properties and desorbing them. On the other hand, by embedding the microcavities in a thick transparent medium such as SiO$_x$ the mode shift with temperature may be suppressed. This might be useful to shift the emission of QDs embedded in microcavities without affecting the cavity properties.
Concerning the tuning of the QD emission, it would be desirable to increase the spatial resolution, so as to be able to independently tune closely-spaced QDs or QDs placed inside the same microcavity~\cite{imamoglu99}. With the approach presented here, different QDs located in the same mesa are observed to shift similarly, since the mesa structure reaches an approximately homogeneous temperature (see Fig. 1D). The spatial resolution might be increased by applying the laser heating, at higher powers than those considered here, to QDs placed on a planar substrate~\cite{brunner92-1}, so that heat can be efficiently dispersed in the surrounding of the illuminated spot. Further improvements may include the use of pulsed lasers and near-field optics~\cite{chimmalgi05}. Using different ILP conditions may also affect the interdiffusion paths and hence the final atomic configurations in a QD.
Since practically all material combinations are sensitive to heat treatment, the approach is obviously not limited to the systems discussed here.

In conclusion, we have demonstrated that {\it in-situ} laser microprocessing can be effectively employed as a post-fabrication tool to engineer the properties of single nanostructures down to the quantum level.

 We acknowledge Ch. Deneke, U. Waizmann, E. Coric, T. Reindl and W. Winter for technical assistance, P. Michler, A. Fiore and M. Jetter for fruitful discussions and F. Horton and M. Benyoucef for contributing to the measurements. This work was financially supported by the BMBF (01BM459) and by the DFG research group ``Positioning of single nanostructures - Single quantum devices''.


\begin{thebibliography}{10}

\bibitem{poate82}
J.~M. Poate, J.~W. Mayer, eds., {\it Laser-Annealing of semiconductors\/}
  (Academic Press, New York, 1982).

\bibitem{allmen95}
M.~V. Allmen, A.~Blatter, eds., {\it Laser-Beam Interaction with Materials;
  Physical Principles and Applications\/} (Springer, Berlin, 1995).

\bibitem{chimmalgi05}
A.~{Chimmalgi}, D.~J. {Hwang}, C.~P. {Grigoropoulos}, {\it Nanoletters\/} {\bf
  5}, 1924 (2005).

\bibitem{baumgartner94}
P.~{Baumgartner}, {\it et~al.\/}, {\it Appl. Phys. Lett.\/} {\bf 64}, 592
  (1994).

\bibitem{deneke04}
C.~{Deneke}, N.-Y. {Jin-Phillipp}, I.~{Loa}, O.~G. {Schmidt}, {\it Appl. Phys.
  Lett.\/} {\bf 84}, 4475 (2004).

\bibitem{brunner92-1}
K.~{Brunner}, G.~{Abstreiter}, M.~{Walther}, G.~{B{\" o}hm}, G.~{Tr{\" a}nkle},
  {\it Surf.\ Sci.\/} {\bf 267}, 218 (1992).

\bibitem{nebel97}
C.~E. {Nebel}, {\it et~al.\/}, {\it J. Appl.\ Phys.\/} {\bf 82}, 1497 (1997).

\bibitem{zaitsev01}
S.~{Zaitsev}, {\it et~al.\/}, {\it Semicond.\ Sci.\ Technol.\/} {\bf 16}, 631
  (2001).

\bibitem{dubowski00}
J.~J. {Dubowski}, C.~N. {Allen}, S.~{Fafard}, {\it Appl. Phys. Lett.\/} {\bf
  77}, 3583 (2000).

\bibitem{djie06}
H.~S. {Djie}, B.~S. {Ooi}, O.~{Gunawan}, {\it Appl. Phys. Lett.\/} {\bf 89},
  1901 (2006).

\bibitem{chimmalgi05-1}
A.~{Chimmalgi}, C.~P. {Grigoropoulos}, K.~{Komvopoulos}, {\it J. Appl.\
  Phys.\/} {\bf 97}, 4319 (2005).

\bibitem{crommie93}
M.~F. {Crommie}, C.~P. {Lutz}, D.~M. {Eigler}, {\it Nature\/} {\bf 363}, 524
  (1993).

\bibitem{vahala03}
K.~J. {Vahala}, {\it Nature\/} {\bf 424}, 839 (2003).

\bibitem{michler03}
P.~Michler, ed., {\it {Single Quantum Dots: Fundamentals, Applications and New
  Concepts}\/} (Springer, Berlin, 2003).

\bibitem{krauss05}
T.~F. {Krauss}, {\it Science\/} {\bf 308}, 1122 (2005).

\bibitem{brozel96}
M.~R. Brozel, G.~E. Stillman, eds., {\it Properties of Gallium Arsenide\/},
  EMIS Datareviews series (INSPEC, London, 1996), third edn.

\bibitem{strauf06}
S.~{Strauf}, {\it et~al.\/}, {\it Appl. Phys. Lett.\/} {\bf 88}, 43116 (2006).

\bibitem{frateschi96}
N.~C. {Frateschi}, A.~F.~J. {Levi}, {\it J. Appl.\ Phys.\/} {\bf 80}, 644
  (1996).

\bibitem{badolato05}
A.~{Badolato}, {\it et~al.\/}, {\it Science\/} {\bf 308}, 1158 (2005).

\bibitem{white05}
I.~M. {White}, N.~M. {Hanumegowda}, H.~{Oveys}, X.~{Fan}, {\it Opt. Express\/}
  {\bf 13}, 10754 (2005).

\bibitem{yoshie04}
T.~Yoshie, {\it et~al.\/}, {\it Nature\/} {\bf 432}, 200 (2004).

\bibitem{reithmaier04}
J.~P. Reithmaier, {\it et~al.\/}, {\it Nature\/} {\bf 432}, 197 (2004).

\bibitem{peter05}
E.~{Peter}, {\it et~al.\/}, {\it Phys.\ Rev.\ Lett.\/} {\bf 95}, 067401 (2005).

\bibitem{michler00}
P.~Michler, {\it et~al.\/}, {\it Science\/} {\bf 290}, 2282 (2000).

\bibitem{stinaff06}
E.~A. {Stinaff}, {\it et~al.\/}, {\it Science\/} {\bf 311}, 636 (2006).

\bibitem{xie05}
Z.~G. {Xie}, G.~S. {Solomon}, {\it Appl. Phys. Lett.\/} {\bf 87}, 3106 (2005).

\bibitem{leon97}
R.~{Leon}, D.~R.~M. {Williams}, J.~{Krueger}, E.~R. {Weber}, M.~R. {Melloch},
  {\it Phys.\ Rev.\ B\/} {\bf 56}, 4336 (1997).

\bibitem{offermans05}
P.~{Offermans}, {\it et~al.\/}, {\it Phys.\ Rev.\ B\/} {\bf 72}, 165332 (2005).

\bibitem{imamoglu99}
A.~{Imamo{\u g}lu}, {\it et~al.\/}, {\it Phys.\ Rev.\ Lett.\/} {\bf 83}, 4204
  (1999).

\end{thebibliography}


\clearpage

\begin{figure}
\epsfig{file=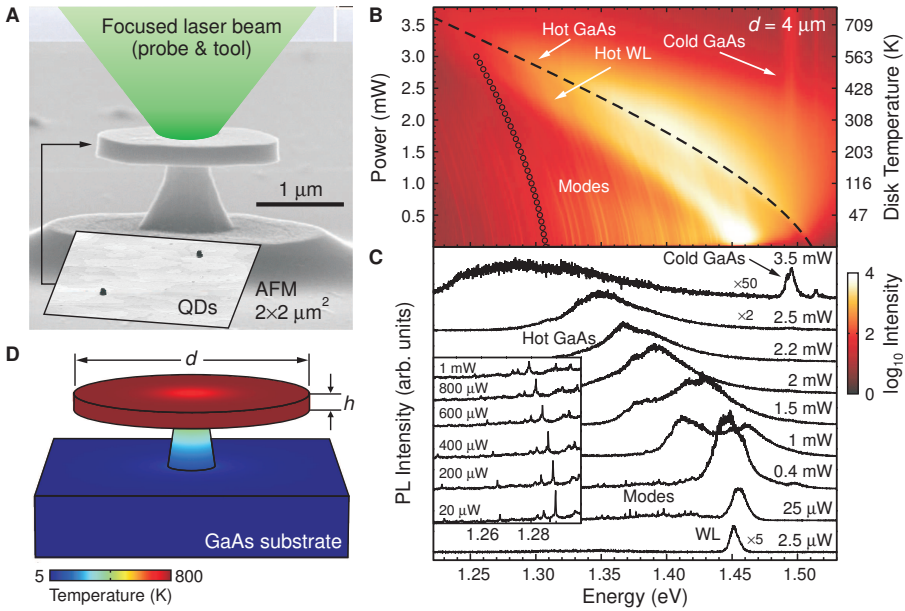, width=12cm}
\caption{(A) Scanning electron microscopy image of a GaAs microdisk containing In(Ga)As QDs and schematic illustration of the {\it in-situ} laser processing (ILP) method. Inset: Atomic force microscopy image of self-assembled In(Ga)As QDs (dark in the image) with low surface density. (B) Photoluminescence (PL) intensity as a function of energy and laser power. (C) Selected spectra showing the red-shift of the emission peaks produced by the laser heating. Inset: close-up spectra illustrating the mode shift. (D) Calculated temperature profile for a disk similar to that used for the measurements shown in (B) and (C).}
\label{fig1}
\end{figure}

\begin{figure}
\epsfig{file=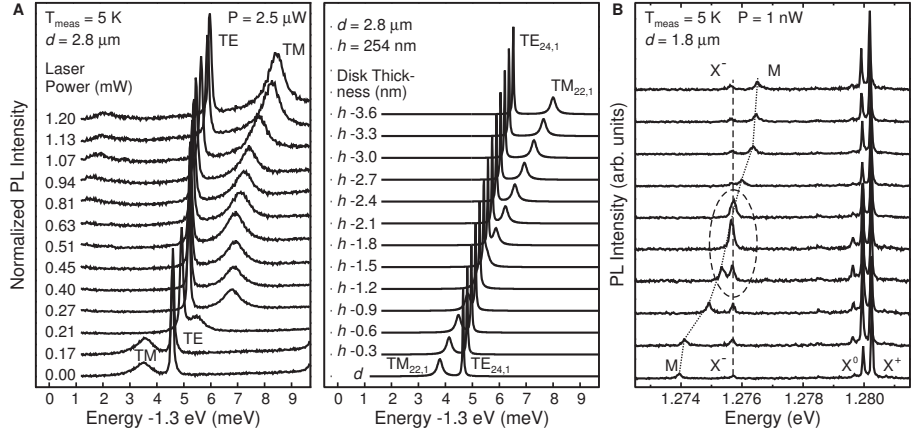, width=12cm}
\caption{(A) PL spectra showing the tuning, by ILP, of TE and TM cavity modes in a microdisk with 2.8~$\mu$m diameter (left panel) and calculated spectra for a 2.8~$\mu$m-diameter disk with decreasing thickness $h$ (right panel). 
In the calculated spectra, the whispering gallery modes are labeled according to their azimuthal and radial numbers~\cite{frateschi96}.
(B) Tuning of a cavity mode $M$ into (and off) resonance with the negative trion X$^-$ confined in a single QD.}
\label{fig2}
\end{figure}

\begin{figure}
\epsfig{file=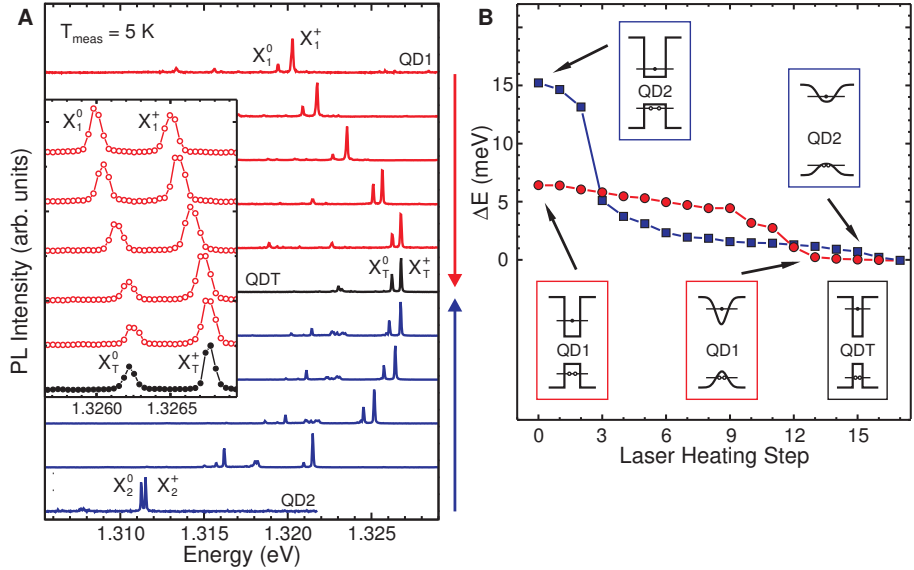, width=12cm}
\caption{(A) PL spectra illustrating the tuning, by ILP, of the positive trion X$^+$ emission of three QDs into resonance. The energy of the X$^+_T$ line of the QD labeled as QDT is taken as target. Inset: PL spectra showing the fine-tuning of the QD1 emission. (B) Shifts of the QDs shown in (A) as a function of heating step. Insets: schematics of the band structure of the QDs prior to and after ILP.}
\label{fig3}
\end{figure}
\end{document}